\documentclass[letter,traditabstract]{aa}

\pdfoutput=1
\usepackage{amsmath}
\usepackage{algorithm}
\usepackage{algpseudocode} 
\usepackage{graphicx}
\usepackage{txfonts}
\usepackage{mathrsfs}
\usepackage{natbib}
\usepackage{color}
\usepackage{soul}
\usepackage{multirow}
\usepackage[multiple]{footmisc}
\usepackage{url}

\def\dv{\boldsymbol{d}}
\def\mv{\boldsymbol{m}}

\begin{document}

\title{Butterfly diagram of a Sun-like star observed using asteroseismology}
\author{
  M.~Bazot\inst{1,2}
  \and M.~B.~Nielsen\inst{2}
  \and D.~Mary\inst{3}
  \and J.~Christensen-Dalsgaard\inst{4}
  \and O.~Benomar\inst{2}
  \and P.~Petit\inst{5}
  \and L.~Gizon\inst{6,7,2}
  \and K.~R.~Sreenivasan\inst{2,8}
  \and T.~R.~White\inst{4}
}
 
\institute{Division of Sciences, New York University Abu Dhabi, United Arab Emirates; mb6215@nyu.edu
\and
Center for Space Science, NYUAD Institute, Abu Dhabi, UAE
\and
Laboratoire Lagrange, Universit\'e C\^ote d'Azur, Observatoire de la C\^ote d'Azur, CNRS, Boulevard de l'Observatoire, CS 34229, 06304 Nice cedex 4, France
\and
Stellar Astrophysics Centre, Department of Physics and Astronomy, Aarhus University, Ny Munkegade 120, DK-8000 Aarhus C, Denmark 
\and
IRAP (Institut de Recherche en Astrophysique et Plane\'etologie), Universit\'e de Toulouse, CNRS, CNES, UPS, F-31400 Toulouse, France
\and
Max-Planck-Institut f\"ur Sonnensystemforschung, G\"ottingen, Germany
\and
Institut f\"ur Astrophysik, Georg-August-Universit\"at G\"ottingen, G\"ottingen, Germany
\and
New York University, NY 10012, USA
}
\abstract
{Stellar magnetic fields are poorly understood but are known to be important for stellar evolution and exoplanet habitability. They drive stellar activity, which is the main observational constraint on theoretical models for magnetic field generation and evolution. Starspots are the main manifestation of the magnetic fields at the stellar surface. In this study we measure the variation of their latitude with time, called a butterfly diagram in the solar case, for the solar analogue HD~173701 (KIC~8006161). To that effect, we use \emph{Kepler} data, to combine starspot rotation rates at different epochs and the asteroseismically determined latitudinal variation of the stellar rotation rates. We observe a clear variation of the latitude of the starspots. It is the first time such a diagram is constructed using asteroseismic data.}

\maketitle

\section{Introduction}

Solar activity is characterized by an 11-year cycle in the number and area of sunspots \citep{Schwabe1844}. Its monitoring is important in many fields, such as Earth climate \citep{Haigh07} or space travel studies \citep{Pulkkinen07}. Sunspots are regions of high concentration of the solar magnetic field \citep{Solanki03}, indicating that the latter is the main driver of the activity cycle. In order to understand the solar magnetic behaviour, dynamo models have been developed \citep{Charbonneau10}. They aim at explaining how an initial weak magnetic field can be amplified to the values observed in the Sun. A traditional observational test to which these models ought to comply is to reproduce the butterfly diagram, which describes the evolution of the latitude of the solar active regions with time \citep{Maunder1904}. Activity is also observed in other stars. It is important to understand stellar evolution, for instance through magnetic braking \citep{Thompson03}, and exoplanet habitability \citep{Vidotto13}.

Recovering stellar butterfly diagrams is however a difficult task that requires to locate individual spots or groups of spots on the stellar surface. Spot mapping using photometric data alone is known to be hampered by degeneracies in light curve models \citep[e.g.][]{Walkowicz13}, so that spectroscopic or interferometric data are usually favoured in order to recover stellar brightness maps \citep{Vogt83, Roettenbacher16}. A number of active targets have been monitored through long-term Doppler mapping campaigns, although due to sparse temporal sampling these time-series have rarely led to actual butterfly diagrams \citep[e.g.][for II~Peg]{Hackman11}. In any case, Doppler mapping or long baseline interferometric imaging can only detect very large stellar spots, and are therefore limited to the most active stars, quite far from the Sun in terms of magnetic properties. We propose here an original method that uses \emph{Kepler} time series to measure the latitudes of the active regions of stars and construct a stellar butterfly diagram. We apply it to the sun-like star HD~173701 (KIC~8006161).

\section{Method}\label{sect:methodo}

Our approach uses \emph{Kepler} \citep{Jenkins10} photometric time series and is divided into two stages. First we wish to obtain information on the large-scale rotational flow in the stellar interior, in particular in the convective envelope below the surface, using the information contained in the oscillation frequencies of the global acoustic pulsation modes (p modes) of the star (Sect.~\ref{sect:rotprof}). To that effect we use the recent results of \citet{Benomar18}. Second, we seek to measure the rotation rates of active regions, carried by the surface rotational flow (Sect.~\ref{sect:rotGP}). Third, we construct the butterfly diagram of HD~173701 over the duration of the \emph{Kepler} mission by inverting the rotation-rate measurements thanks to the asteroseismically derived rotation profile (Sect.~\ref{sect:papillon}).

We note that, in Sun-like stars, these measurements are uncorrelated since the observed pulsation modes of the stars are only affected by large-scale flows. In the frequency space, these two contributions are well separated, the modes having characteristic frequencies of the order of a few thousand $\mu$Hz, while the rotation rates inferred from the starspots are of the order of a few hundred nHz.

\subsection{Asteroseismic rotation profile}\label{sect:rotprof}

Our starting point is the derivation of a rotational profile of a Sun-like star. It is well established that the solar radiative interior rotates as a solid body while the rotation rate of its outer convective envelope varies with latitude \citep{Thompson03}. For the Sun, the equator rotates faster ($\sim$476~nHz) than the pole ($\sim$320~nHz). Such differential rotation is most probably the result of an interplay between rotation and convection in the solar envelope \citep{Ruediger74,Gilman81}. 

Rotation affects stellar pulsations. The classical framework to describe stellar oscillations is built on the perturbed equations for the static stellar structure \citep{Aerts10}. It can be shown \citep{LB67} that superimposing a small-amplitude rotational velocity field on the hydrostatic stellar structure will lift a degeneracy of the eigenfrequencies of the p modes, an effect usually termed {\textquotedblleft}rotational splitting{\textquotedblright}. The degenerate frequency then becomes a multiplet whose distribution depends on the solid-body component of rotation rate of the star and on the magnitude of the differential rotation in the convective zone. If one defines a theoretical model $\Omega = \Omega(\theta)$ for the latitudinal differential rotation, with $\theta$ being the co-latitude, it is possible to relate it to the frequency splitting. This means that one can infer a differential-rotation profile if there exist precise enough measurements of the frequency splitting.

In a recent work \citep{Benomar18}, contributions from latitudinal differential rotation to the above mentioned rotational splitting were measured for a set of Sun-like stars. It includes the solar analogue HD 173701, which has a mass $M = 0.95$~M$_{\odot}$, with M$_{\odot}$ the solar mass, and an age $t_{\star} = 4.49$~Gyr \citep{SA17}. This star rotates on average 1.33 times faster than the Sun, with a measured bulk rotation rate of 566~nHz. It has been observed for approximately four years by \emph{Kepler}. Detailed modelling of its acoustic power spectrum led to a significant detection of non-zero latitudinal differential rotation. In the following we explain how to extend such results with rotation-rate measurements in order to construct the butterfly diagram of the HD~173701. For the sake of completeness we review the work of \citet{Benomar18} in more details in Appendix~\ref{app:astero}.

At this point we have to make a further assumption relative to \citet{Benomar18}. It has been shown by \citet{Kiefer17}, \citet{Santos18}, \citet{Salabert18} that the oscillation frequencies of HD~173701 shift with the activity cycle. We therefore need to take these shifts to be constant over frequency ranges of the order of a rotational splitting.

\subsection{Photometric characterization of active regions}\label{sect:rotGP}

\begin{figure}[h!]
\center
\includegraphics[width=\columnwidth]{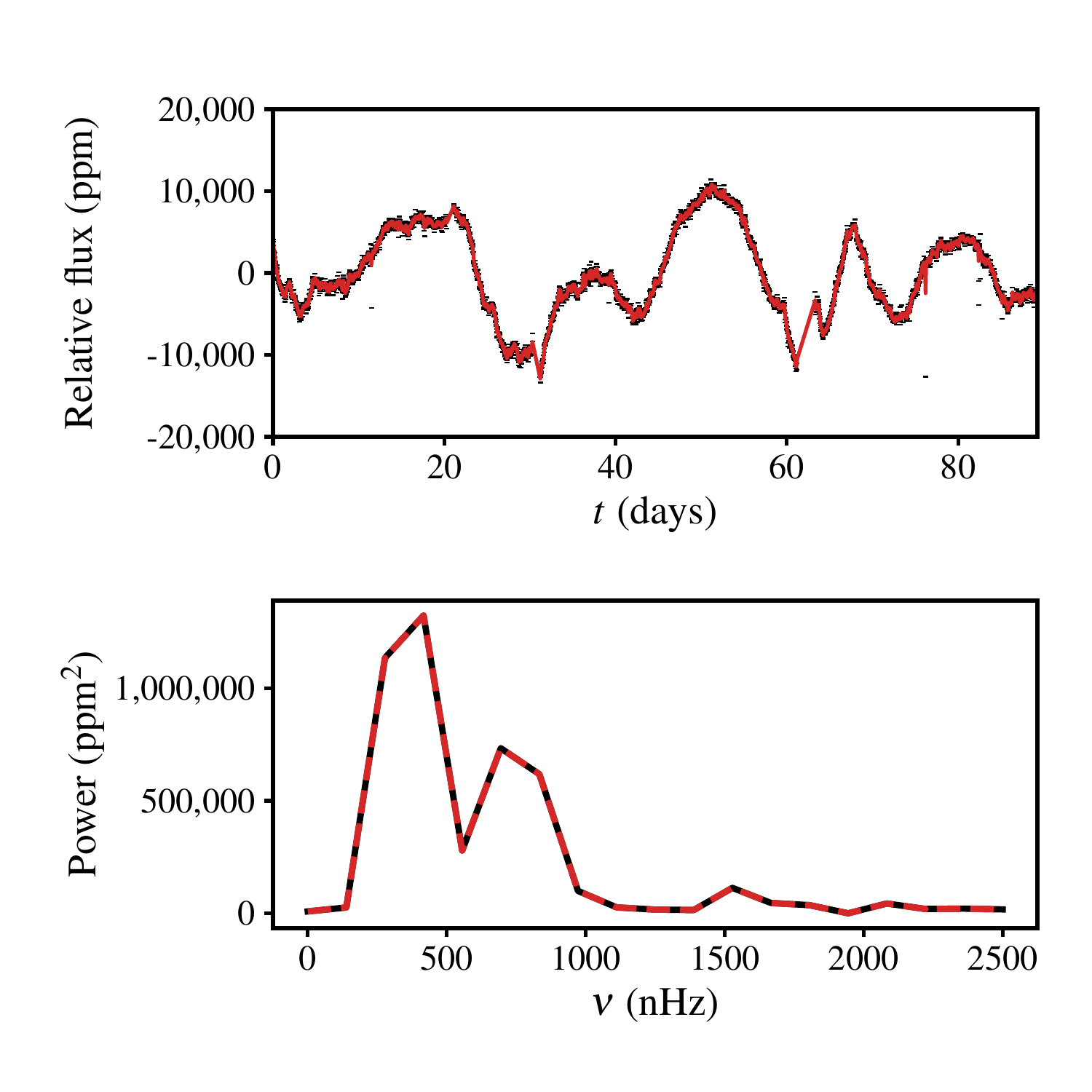}
\includegraphics[width=\columnwidth]{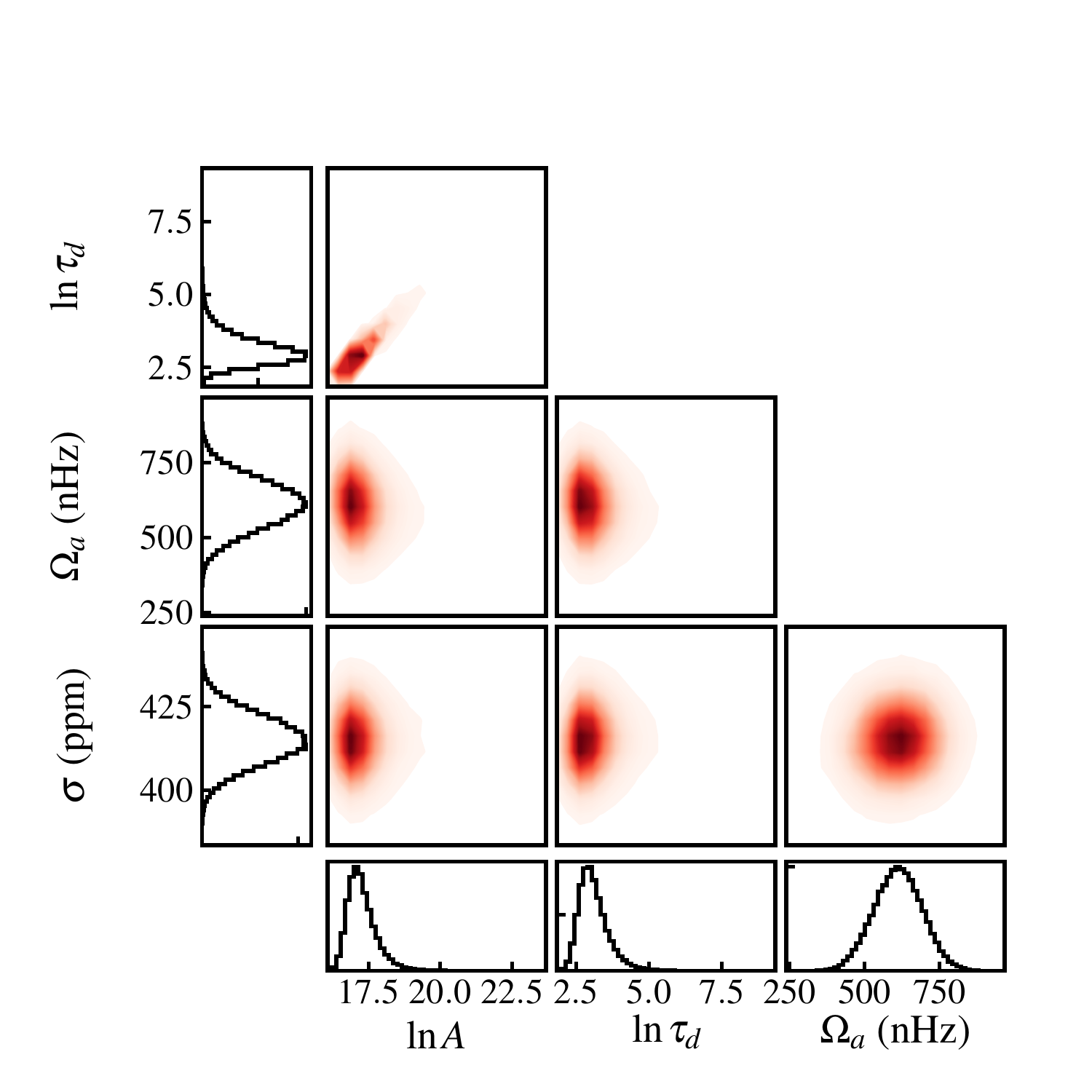}
\caption{Upper panel: Time series for the relative flux HD~173701 during Q3. The black dots mark the measurements and associated errors. The red line shows the inferred mean value of the posterior predictive density conditional on the inferred MAP of the parameters \citep{Gelman04}. Middle panel: Corresponding power spectrum computed using the Lomb-Scargle periodogram \cite{Scargle82}. The black line represents the observed power spectrum. The red line is the power spectrum of the mean value of the conditional posterior predictive density. Lower panel: Marginalized 1-dimensional and 2-dimensional PDFs for the parameters of the correlation function of the Gaussian Process used to model the time series.}
\label{fig:GP}
\end{figure}

\begin{figure*}
\includegraphics[width=\textwidth]{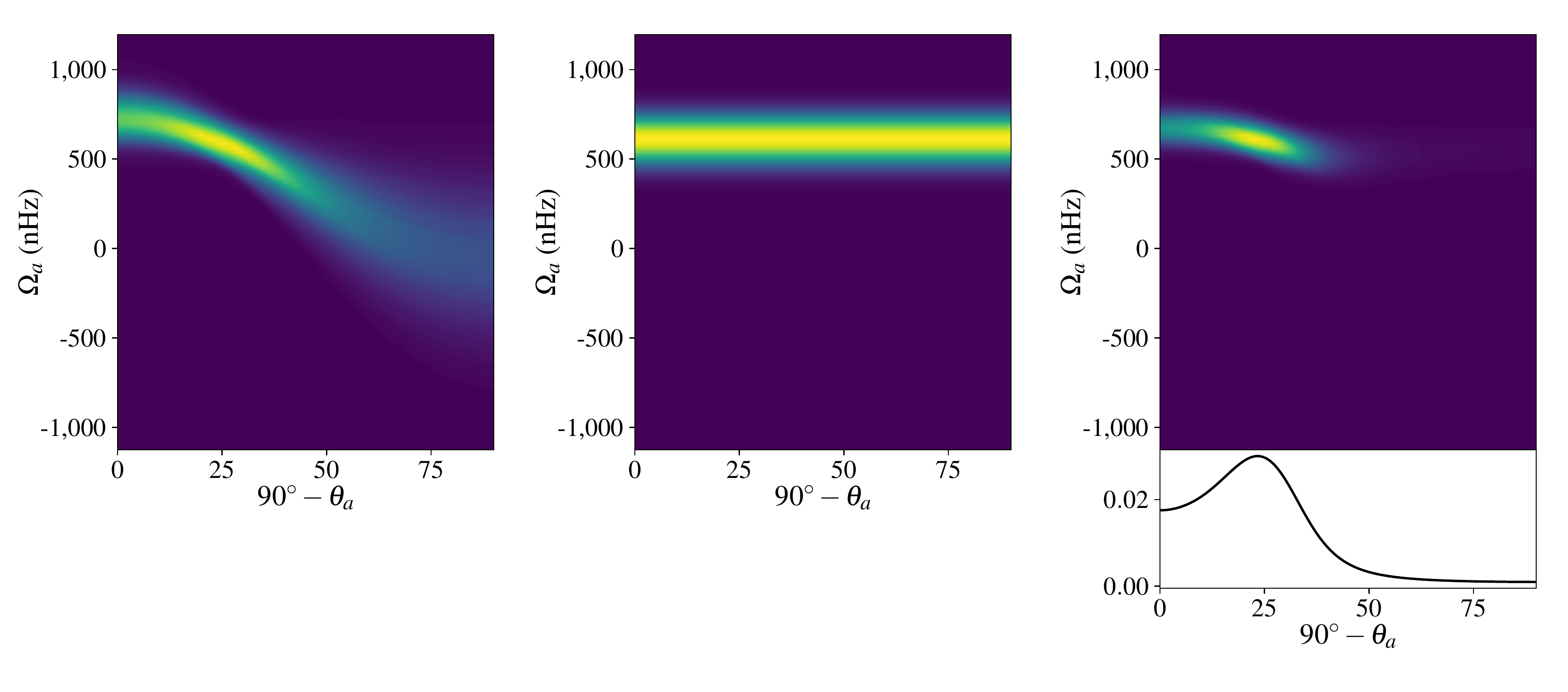}
\caption{Probability density for the latitude of an active region at median time of quarter Q3. The left panel shows the theoretical density for the couple $(\Omega_{a},\theta_{a})$. The central gives the prior density for the couple $(\Omega_{a},\theta_{a})$, it is Gaussian for $\Omega_a$ and uniform for $\theta_a$. The upper right panel gives the posterior density for the parameters. The lower right panel is the marginal density for $\theta_a$, obtained after integration over $\Omega_a$.}
\label{fig:densities}
\end{figure*}

Once obtained, a relation $\Omega(\theta)$ can be used to infer the co-latitude, $\theta_a$, of a stellar active region such as a group of spots provided its rotation rate, $\Omega_a = \Omega(\theta_a)$, can be determined. The method for rotation rate measurements is inspired by solar observations and uses photometry. The Total Solar Irradiance varies during the activity cycle due to bright (plages) and dark (spots) regions, moving across the solar disk \citep{Frohlich04}. A similar behaviour is observed for other stars using photometric measurements. During its lifetime, an active region will produce a quasi-periodic signal in the photometric time series. The long-cadence \emph{Kepler} time series allow us to measure such modulations for HD~173701.

To measure the period of this modulation, we model the time series using Gaussian processes. Owing to the intermittent influence of the plages and spots, and to the stellar limb darkening, this signal departs significantly from pure harmonic oscillations. Besides, the noise impacting the photometric times series has a frequency-dependent part mainly caused by surface granulation. The time series model depends on several parameters, which are estimated using an MCMC algorithm. These parameters are the amplitude of the modulation induced by the active region, $A$ (in ppm), the lifetime of the active regions $\tau_d$ (in days), the rotation rate $\Omega_a$ (in nHz) and an additional noise component, $\sigma$ (in ppm), describing residuals not captured by the other component of the Gaussian process (see Appendix~\ref{app:rotation} for details). 

We model independently the time series for each \emph{Kepler} quarter which span from Q0 to Q17. The fit obtained for Q3, representative of our results, is shown in Fig.~\ref{fig:GP}; it reproduces well the time series and its frequency spectrum. Our analysis is restricted to a region in the frequency spectrum surrounding the low-frequency activity peak (Fig.~\ref{fig:GP}, middle panel). In fact we observe multiple peaks potentially corresponding to several active regions. We consider that each of these peaks correspond to a spot or a group of spots that rotates at the same latitude. We treat them as a single active region and consider the rotation rate of its barycentre. This is motivated by the fact that, when counting sunspots, a larger weight is given to groups \citep{Hathaway10}. The independent modelling, quarter by quarter, allows us to measure the evolution of the rotation rate with time. Note that quarters Q0, Q1 and Q17 were left out because the corresponding time series are to short to obtain robust results.


The Gaussian processes allow to reproduce very well the temporal variation of the intensity curve as can be seen in the upper panel of Fig~\ref{fig:GP}. Using the marginal density of the rotation rate $\Omega_a$, we can estimate its Maximum A Posteriori (MAP). Such a density is seen for Q3 in the lower panel Fig~\ref{fig:GP}, corresponding to a MAP of 704~nHz.  The  MAP estimates of the rotation rates for all quarters are in the range 330 -- 985~nHz, which already indicates that the spots are migrating along the stellar surface. In general the amplitudes of the signal and the lifetimes of the active regions are poorly constrained. Since they are strongly correlated, a wide range of values for the lifetime and the amplitude can lead to good models. Consequently, their estimated values cover several orders of magnitude. This does not impact the final result since these parameters are uncorrelated from the others.

\subsection{Inversion for the latitude of the active regions}\label{sect:papillon}

Our goal is now to invert for the latitude of the active region at each for each quarter we selected. Since the latitudinal rotation profile has been derived using the entire \emph{Kepler} time series, it is necessary to assume that the properties of differential rotation do not vary on time scales comparable to the activity cycle. In the solar case, this is verified to a very good approximation \citep{Thompson03}. 

In order to invert for the latitude of the active regions, we need to take into account all the errors that may affect our final estimate of $\theta_a(t)$. These are of two kinds. The first one is the error on the measurement of the rotation rate, the second is the uncertainty in the theoretical model for the rotation rate, $\Omega(\theta)$, as obtained by \citet[][see also Appendix~\ref{app:astero}]{Benomar18}.

A generic framework to solve an inverse problem that takes into account these two sources of error is given by the concept of conjunction of states of information \citep[][see Appendix~\ref{app:papillon}]{Tarantola82}. In the following we use the more compact notation $\Omega_a(t_i)\coloneqq \Omega_i$ and $\theta(t_i) \coloneqq \theta_i$, with $i = 1,\dots,N$ and $N$ the number of rotation rate measurements. The posterior density for $(\Omega_i,\theta_i)$, which represents the state of information (or state of knowledge, see Appendix~\ref{app:papillon}) we have once all the errors have been considered, can be written
\begin{equation}\label{eq:soi}
\displaystyle
\sigma_i(\Omega_i,\theta_i) = \frac{\rho_i(\Omega_i,\theta_i)\Theta(\Omega_i,\theta_i)}{\mu(\Omega_i,\theta_i)}.
\end{equation}

\begin{figure*}
\center
\includegraphics[width=0.9\textwidth]{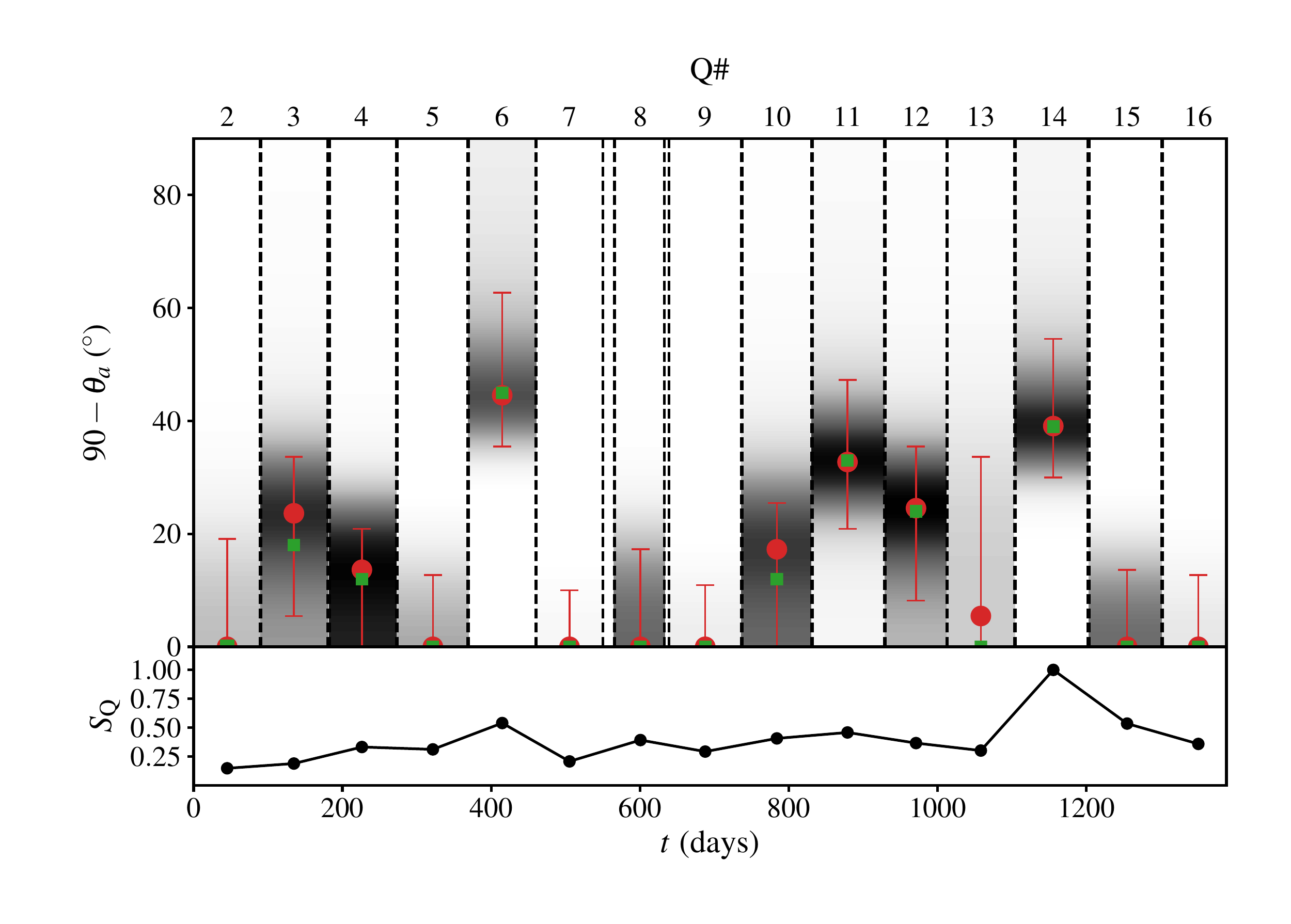}
\caption{Asteroseismic butterfly diagram of HD~173701, giving the latitude of the active region as a function of the median time of each quarter (labelled on the upper axis). The grey shades show the posterior densities obtained for each of the selected quarters and normalized to the global maximum; greyscale from 0 (white) to 1 (black). Red dots mark the Maximum A Posteriori estimates of the latitude. The vertical bars give the limits of the corresponding 68.3\% credible intervals. The green squares show the maximum of the density of the 1024 estimates obtained from artificial time series. Lower panel: Variation of amplitude of the signal, $S_Q$, normalized to its maximum.}
\label{fig:papillon}
\end{figure*}

The theoretical density $\Theta(\Omega_i,\theta_i)$ describes the error on the model. It is basically the relation $\Omega_i = \Omega_i(\theta_i)$ with an associated error bar. The prior density $\rho_i(\Omega_i,\theta_i)$ is the combination of the information obtained from the observation on $\Omega_i$ (see for instance Fig.~\ref{fig:GP}, lower panel) and the prior information on the parameter $\theta_i$ (which has been chosen uniform over the range $0^{\circ} - 90^{\circ}$). The posterior density $\sigma_i(\Omega_i,\theta_i) $ results from the conjunction of these two pieces of information. In Fig.~\ref{fig:densities} we show these three densities for quarter Q3, respectively in the left, central and right panel. The marginal posterior on $\theta_i$ is shown in the lower right panel, and is obtained from the integration over all possible rotation rates of the joint posterior. The meaning of the homogeneous density $\mu(\Omega_i,\theta_i)$ is given in Appendix~\ref{app:papillon}.

\section{Result}\label{sect:result}

We use the marginal posteriors of $\theta_a(t)$ to derive the butterfly diagram pictured in Fig.~\ref{fig:papillon} for all selected quarters. The latitude of the active region is restricted to the range $[0^{\circ} -90^{\circ}]$ since we do not resolve the stellar disk and thus do not know on which hemisphere is located the spot.

The credible intervals were computed as the smallest intervals that encompass the MAP and over which the probability mass is 0.683. There is a clear variation in the data. Half the active regions are found at the equator. Among the other half, five clearly exclude the $0^{\circ}$ latitude at a 68.3\% level. The active regions seen in Q6 and Q14 exclude it at a 99.7\% level. It is thus ruled out that the signal seen could be due to equatorial active regions, with the high-latitude ones being outliers driven by intrinsic stochastic variability.


The estimated co-latitude (red dots in Fig.~\ref{fig:papillon}) rely on the parameters of the Gaussian process, which were estimated from the time series. This means that different data sets of the same star in the same activity configuration would lead to different inferred latitudes. In order to investigate the variability of the angle estimation procedure (and hence check for the robustness of these results), we made Monte Carlo simulations. We simulated 1024 time series using the MAP estimates of the parameters of the Gaussian Processes for each quarter\footnote{To generate samples we used the predictive probability density \citep[][Sect~2.1.1]{Rasmussen05}.}. We applied the latitude-estimation process to each artificial time series, resulting in a sample of 1024 MAPs estimates for each such latitude of the active region. The distribution of this MAPs obtained from artificial time series can be compared to the MAP obtained directly from the actual data. The location of the maxima for these distributions are shown in Fig.~\ref{fig:papillon} as green squares. The MAP locations are essentially not impacted by the noise realization, which supports the fact that the co-latitude extracted from the \emph{Kepler} data for this star must be close to the values shown in Fig.~\ref{fig:papillon}. Slightly larger deviations are seen when the uncertainty on $\theta_a$ is large, which is to be expected. 


Over the four years of measurement, the active regions remain located at latitudes below $50^{\circ}$. The butterfly diagram clearly shows an alternance between equatorial and high-latitude active regions. 

In the lower panel of Fig.~\ref{fig:papillon} we plotted $S_Q$, which we deifne as the standard deviation of the light curve evaluated over quarter $Q$. We used this quantity as a proxy for the activity index $S_{\mathrm{ph}}$ rather than the definition given in \citet{Garcia14} for two reasons. First it involves an evaluation over a period that is five times the rotation rate, which becomes hard to interpret when we take into account the variability of the rotation rate. Second, if we retain the average rotation rate of KIC~8006161, then a quarter is approximately 4.5 times larger, which is commensurate with the definition of $S_{\mathrm{ph}}$ and preserves the statistical independence of the points in our time series. 

  We first see that low-amplitude modulations are observed for active regions at high latitude and near the equator. Let us assume that the amplitude of the signal follows the variations of the fraction of the stellar surface covered by the active regions. Then this observation in qualitative agreement with what is already known for the Sun \citep[][ Fig.~28]{Hathaway10} where small spots are observed at all latitudes (within the band in which they are confined). The two highest-amplitude modulations are seen for the regions with the highest latitudes, during Q6 and Q14. If our assumption that the amplitude relates to the surface covered by the active regions, then this departs from what is seen in the solar case. There the largest spots are seen at mid-latitudes.

Extracting a periodicity for the activity cycle from these measurements remains a difficult task. The periodogram of the signal does not show any significant peak at low frequencies. Time series with longer spans would be needed to obtain such an estimate. It is interesting to note that the two highest-latitude spots also correspond to the longest estimated lifetimes (in the sense of the MAP), with $\tau_d \gtrsim 1$~yr. This is of course an overestimation, since the signal lasts less than this characteristic time (see Appendix~\ref{app:rotation}). It should be noted that the lifetime of the active region and its rotation rate are uncorrelated, we can therefore safely assume that these biases do not impact the estimates of the latitude for these epochs (as can be checked in Fig.~\ref{fig:GP} for quarter Q3). Such values nevertheless indicate that active regions are producing a coherent signal over a large fraction of quarters Q6 and Q14. This could correspond to one or several long-lived large spots. Again this seems to depart from what is known in the Sun, for which spot lifetimes are much smaller. 

Other activity measurements have been obtained for HD~173701. \citet{Karoff18} have obtained Ca II H and K line measurements that can track the activity cycle. \citet{Kiefer17}, \citet{Santos18} and \citet{Salabert18} have all measured consistent values for the activity-induced frequency shifts over the last eleven quarters of the of the Kepler mission. These latter measurements correlate well with $S_{\mathrm{ph}}$. Further work and larger data sets are needed to understand precisely how the butterfly diagram correlates with these other indicators.

\section{Conclusion}

In this letter we presented the first stellar butterfly diagram derived obtained by combining information inferred from asteroseismic and photometric analyses. Provided an approximate stellar model is known, the only data required to perform the inversion is a photometric times series (collected  from \emph{Kepler} in this study). This approach identifies a powerful link between asteroseismology and other branches of stellar physics studying stellar magnetism, for instance Zeeman-Doppler Imaging \citep{Semel89}. Stellar butterfly diagrams, out of reach of Zeeman-Doppler mapping when it is applied to solar analogues, nicely complement the long-term monitoring of large-scale magnetic geometries still accessible for low activity stars. Both proxies of stellar cycles offer complementary views to understand the underlying dynamo processes. The technique itself requires only moderate computational time and can be envisaged as a systematic processing for surveys of star spots in the perspective of forthcoming the space mission PLATO.

\begin{acknowledgements}
This material is based upon work supported by the NYU Abu Dhabi Institute under grant G1502. Funding for the Stellar Astrophysics Centre is provided by The Danish National Research Foundation (Grant DNRF106). In memory of Albert Tarantola.
\end{acknowledgements}

\bibliography{papillon_ref}

\appendix

\section{Inference of the differential rotation profile}\label{app:astero}

In this section we describe briefly the principles of asteroseismic inversion that allow us to estimate the parameters of our model for the stellar rotation rate $\Omega(\theta)$. In generic form it can be written as an expansion over a basis of functions $W_s(\theta)$ depending on even powers of $\cos(\theta)$ \citep{Brown89}
  \begin{equation}\label{eq:rrate}
    \displaystyle
    \Omega(r,\theta)  = \sum_{s = 0}^{s_{\mathrm{max}}} \Omega_s(r) W_s(\theta).
  \end{equation}
  The functions $\Omega_s(r)$ are often chosen as piecewise continuous functions in order to account for the change in rotational regime between the radiative interior and the convective envelope. In the case of HD~173701, the $\Omega_s(r)$ are chosen constants, and these are the parameters we estimate.
  
The rotational splitting can also be expressed as a basis expansion of the form \citep{Brown89}
  \begin{equation}\label{eq:splitting}
    \displaystyle
    \delta\nu_{n,l,m}  = \sum_{j = 0}^{j_{\mathrm{max}}} a_{2j+1}(n,l) \zeta_j^{(l)}(m),
  \end{equation}
  where the $\zeta_j^{(l)}(m)$ form an orthogonal basis obeying $\sum_{m}  \zeta_i^{(l)}(m)\zeta_j^{(l)}(m) = 0$ if $i\neq j$. Finally, the $a_j$ and $\Omega_s$ are related through the integral equation
\begin{equation}\label{eq:splitkernel}
  \displaystyle
  \delta\nu_{n,l,m} = \int_0^{\pi}\int_0^{R_{\star}}K_{n,l,m}(r,\theta)\Omega(r,\theta)rd\theta dr,
\end{equation}
where $K_{n,l,m}(r,\theta)$ is a kernel that depends on the equilibrium stellar structure and the eigenfunction of the corresponding p modes \citep{Hansen77}. To obtain these kernels we numerically solved the equations for the stellar structure and pulsations, using, respectively, ASTEC \citep{JCD08a} and {\tt adipls} \citep{JCD08b}. 

\begin{figure}[h]
\includegraphics[width=\columnwidth]{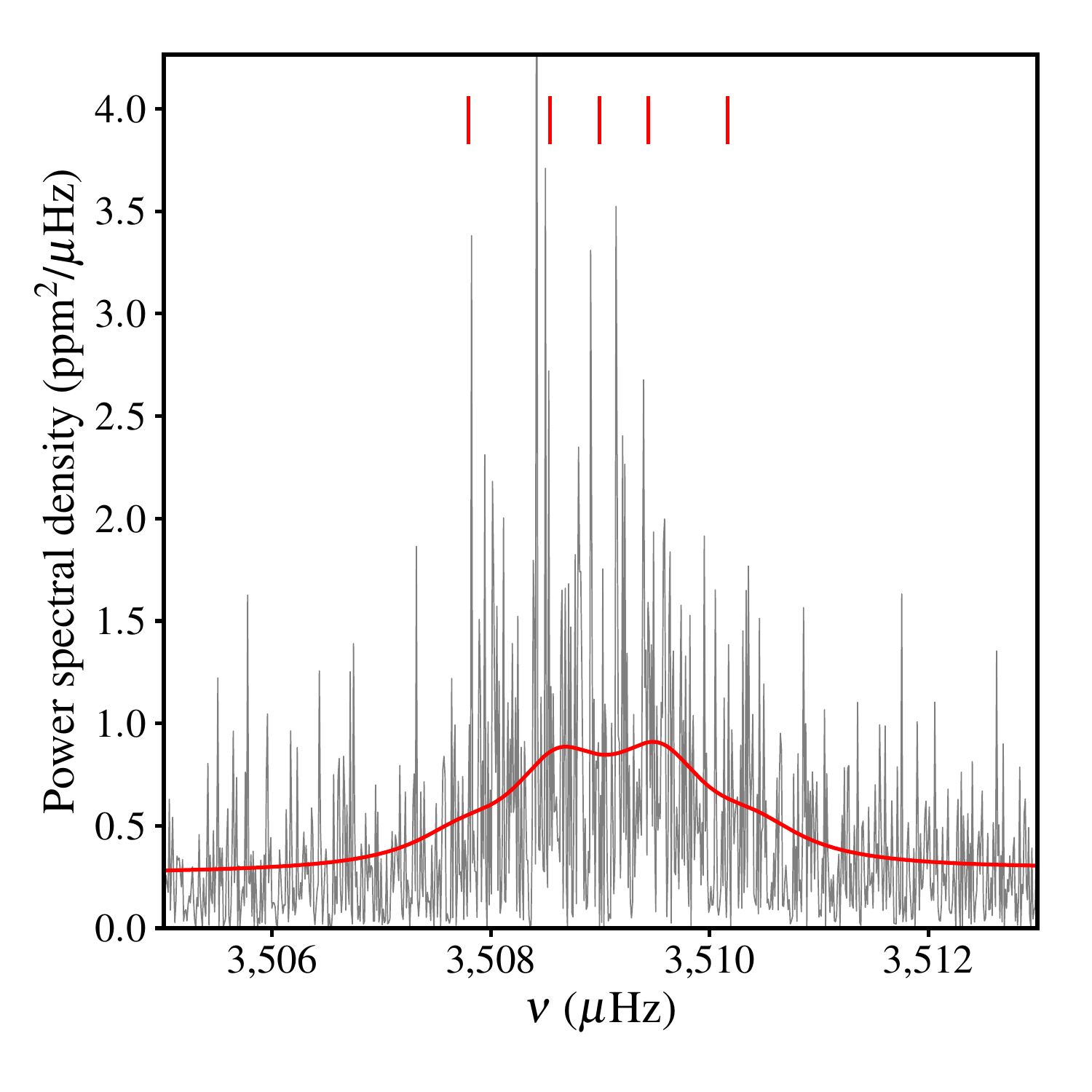}
\includegraphics[width=\columnwidth]{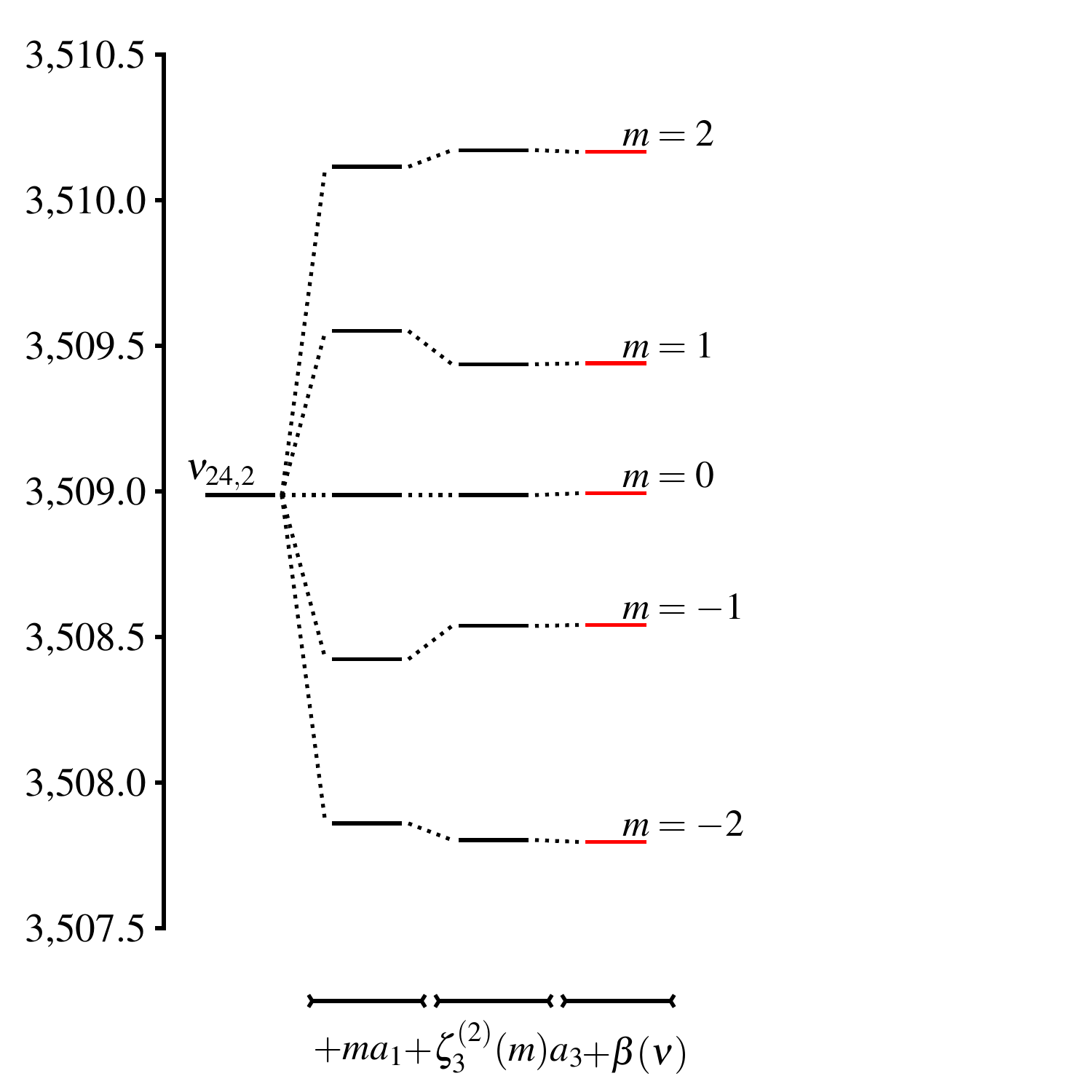}
\caption{Top panel: Power spectrum of the short-cadence \emph{Kepler} data in the region of the eigenfrequency multiplet centred around the frequency $\nu_{n=24,l=2,m=0}$. The black and red lines show respectively the data and the best-fit model of the power spectrum. The vertical red ticks mark the frequencies of the eigenmodes of the multiplet, for $-2 \leq m \leq 2 $. Bottom panel: Splitting diagram for the multiplet. The first two terms of the splitting sequence correspond to the rotational effects while $\beta(\nu)$ is an additional term that describes aspherical contributions to the eigenfrequencies. The red horizontal ticks correspond to those seen in the bottom panel.}
\label{fig:splitting}
\end{figure}

A judicious choice of the basis functions $\zeta_j^{(l)}$ will ensure that there exists a one-to-one relation between the coefficients $a_{2s+1}$ of the splitting expansion and the $\Omega_s$ of the rotation rate expansion \citep{Ritzwoller91,Schou94}. The resulting orthogonality condition on the $W_s$ allows to derive their functional form \citep{Pijpers97}, which is $W_s(\theta) = P_{2s+1}^1(\cos\theta)/\sin\theta$, with $P_l^m(\cos\theta)$ being the associated Legendre polynomial of degree $l$ and order $m$. Using this expression and setting $s_{\mathrm{max}} = 1$ in (\ref{eq:rrate}), we obtain the formula used for the rotation rate in this study

\begin{equation}\label{eq:omegatheta}
\Omega(\theta) = \Omega_0 - 1.5\Omega_1(5 \cos^2\theta - 1)  
\end{equation}

Note that the condition on $s_{\mathrm{max}}$ is justified by the above mentioned one-to-one relation and the fact that the current seismic data only allow us to observe $a_1$ and $a_3$ (tests to detect $a_5$ splitting coefficients in the data considered in this study were inconclusive). The goal is to obtain estimates of $\Omega_0$ and $\Omega_1$ in order to derive the theoretical density of the couple $(\Omega_i,\theta_i)$.

   In practice, the coefficients $a_1$ and $a_3$ were estimated using a parametric model for the power spectrum \citep{Gizon03}, namely a sum of Lorentzian components and some noise. The locations of the Lorentzians are the mode eigenfrequencies. Their distribution is given by the relation
\begin{equation}\label{eq:egnfrq}
  \displaystyle
   \nu_{n,l,m} = \nu_{n,l,0} + \delta\nu_{n,l,m} + \beta(\nu).
\end{equation}
Here $\delta\nu_{n,l,m}$ describes the rotational effects and is given by Eq.~(\ref{eq:splitting}). The additional term $\beta(\nu)$ has been introduced to account for the effects of departures from strict sphericity of the star (centrifugal force, magnetic fields, tidal distortion,\dots) and is a linear function of the central frequency of the multiplet \citep{Gough90}. To estimate $a_3$, mode of degree $l=2$ were used.

The existence of a one-to-one relation between $(a_1, a_3)$ and $(\Omega_0, \Omega_1)$ was then used to estimate these latter. The measured rotational splitting in the observed acoustic spectrum was modelled \citep{Benomar18} using Eq.~(\ref{eq:splitting}) with $j_{\mathrm{max}} = s_{\mathrm{max}} = 1$. The splitting coefficients were considered as free parameters to be estimated. The inferred values are $a_1 = 563\pm69$~nHz and $a_3 = 28.61\pm12.41$~nHz. An example of the modelled power spectrum is given in Fig.~\ref{fig:splitting}, alongside the measured splittings. A model for the rotation rate of the convective zone of the form (\ref{eq:omegatheta}) was used together with kernels computed from stellar models \citep{JCD08a,JCD08b,Lund14} to invert this model.  Estimates for the rotation coefficients $\Omega_0 = 566 \pm 69$~nHz and $\Omega_1 = -104 \pm 45$~nHz were inferred. 

 Finally, we expand slightly on the assumption made concerning the influence of the activity-induced frequency shifts on the estimated value for $a_3$. The assumption we have to make is that the entire multiplet caused by rotational splitting is shifted as a block and thus the frequency splitting remains the same. This allows the activity-induced frequency shifts to depend on the frequency itself, but this variation has to remain negligible at scales of the order of $\sim$3~$\mu$Hz, which is characteristic of a rotational splitting in KIC~8006161. If this holds, then we may expect $a_3$ to be relatively constant with time. Note that this is related but subtly different from the assumption of non-varying differential rotation magnitude with time made at the beginning of Sect. 2.3. This latter says that the magnitude of differential rotation does not change over time. The new assumption says that the measurement of this constant quantity $a_3$ is not biased by time-varying effects affecting the frequencies.

\section{Measuring rotation from active regions}\label{app:rotation}

One element needed to infer the latitude of an active region is a measurement of its rotation rate. It is critical to obtain an estimate of the associated uncertainties in order to compute the conjunction of information states described in Sect.~\ref{sect:methodo} (see also Appendix~\ref{app:papillon}).

  The idea retained here is to fit the low-frequency components of the time series using a Gaussian process. This method has been applied recently to \emph{Kepler} data with some success \citep{Angus18}. The critical point that allowed the use of Gaussian processes to model long time series was the development of methods for the fast inversion of covariance matrices\footnote{We used the {\tt celerite} package for python. \url{https://github.com/dfm/celerite}} \citep{FM17}.

The idea of fitting a Gaussian Process to a time series with $N$ points is to consider it as the realization of a random vector of dimension $N$. In principle one would then have to estimate the mean and the covariance matrix of the parent distribution, i.e. $N(N+3)/2$ parameters. However, the assumption is made that the process is stationary and thus that any given term of the covariance matrix can be determined from a correlation function, $k$, that only depends on the time difference, $k(t_i,t_j) = k(\tau_{ij})$, with $\tau_{ij} = |t_j - t_i|$. Adopting models elsewhere developed \citep{FM17} we chose a function of the form 
    \begin{equation}
      \displaystyle
    k(\tau_{ij})  = \frac{A}{2} e^{ -\tau_{ij}/\tau_d } [ \cos ( 2\pi\Omega_{a}\tau_{ij} ) + 1 ] + \sigma^2\delta_{ij}.
    \end{equation}
    The cosine term on the right-hand side models the rotation rate. The second term including $\sigma^2$ is sometimes dubbed {\textquotedblleft}jitter{\textquotedblright} and may capture potential model errors or compensate for underestimated observational errors \citep{Angus18}.

    The parameters of the covariance model are $\boldsymbol{\lambda} \coloneqq (A,\tau_d,\Omega_a,\sigma)$. We used a Bayesian statistical model to describe each time series. The posterior density of $\boldsymbol{\lambda}$ was estimated using an MCMC algorithm \citep{FM13}. An example of the resulting sampling is shown in Fig.~\ref{fig:GP}. The likelihood is given by the density of the Gaussian process. The components of $\boldsymbol{\lambda}$ were assumed independent and the prior could be written as a product of uniform univariate densities. After several tests we found that we could efficiently sample the posterior density if we fit $A$ and $\tau_d$ in logarithmic space and $\Omega_a$ and $\sigma$ in linear space. We chose uniform priors for $\Omega_a$ and $\sigma$ and log-uniform priors for $A$ and $\tau_d$. The adopted boundaries on $\ln A$ and $\ln \tau_d$ are respectively $[-\infty,+\infty]$ and $[-10, 12]$. For $\Omega_a$ and $\sigma$ they are $[150, 1600]$, $[150, 4500]$. We note that the prior on $\Omega_a$ was set so that our results do not get contaminated by very-low or very-high frequency signals that are not properly reproduced by our model (hence relating to filtering practices that can be found in other studies \citep{Angus18}). These boundaries were set after several tests and trials.

    The prior on the lifetime also demands some caution. It should be noted that its upper boundary is far greater than the duration of any time series we are using in this study. And indeed, some estimates of $\tau_d$ from the MCMC simulation happen to be greater than a quarter duration. This can be viewed in several ways. First, one has to remember that we are fitting a stochastic process to the time series. Obtaining long lifetimes out of the estimation process just means that high values of $\tau_d$ are compatible with the degree of correlation between the different timescales probed by the time series and that this degree always remains high. Another way to look at this is that the stationarity assumption under which we are working breaks down if we try to only model one active region over long series. Consequently, when we encounter large estimates for the lifetimes of the active regions, the precise value of the estimate should be seen as meaningless, or at least severely biased. What is important in those case, is that $\tau_d$ is larger than the duration of the time series, indicating an active region that is stable over the duration of the entire subsample. 

\subsection{Conjunction of information states}\label{app:papillon}

The method of inversion is based on the framework described by \citet{Tarantola82}. In this section we just summarize the main points of the approach and refer to the original paper for further details. The basic postulate is that any state of knowledge on the values of a set of parameters can be described using a density function\footnote{By {\textquotedblleft}state of knowledge{\textquotedblright}, it is meant knowledge on the quantitative characteristics of a system. Therefore the existing information we have on this characteristic can be cast into a density function. This method is not fit to deal with qualitative knowledge one might have on a system.}\footnote{This density is a probability density only if it can be normalized to one.}. The approach of the conjunction of states of information consists in expressing the posterior information on a couple $({\bf d},{\bf m})$, with ${\bf d}$ and ${\bf m}$ being the data and model-parameter vectors, using density functions

\begin{equation}\label{eq:conjunction}
\displaystyle
\sigma(\dv,\mv) = \frac{\rho(\dv,\mv)\Theta(\dv,\mv)}{\mu(\dv,\mv)},
\end{equation}
where the densities $\sigma$, $\rho$, $\Theta$ and $\mu$ represent states of information. They are not necessarily probability densities, which would require that they be normalizable. $\sigma$ represents the conjunction of the two states of information described by $\rho$ and $\Theta$. The density $\mu(\dv,\mv)$ is sometimes called a homogeneous probability density and represents the state of null information \citep{Tarantola04}. This means that $\mu$ is the density that is the least informative\footnote{As explained below, we are able to choose $\mu$ constant. However, the precise choice of the homogeneous density is a delicate problem that has been discussed for instance by \citet{Jaynes68}.} on the values of the couple $(\dv,\mv)$. It is present in Eq.~(\ref{eq:conjunction}) so that the conjunction of any state of information by the null state results in no loss of information. The functions $\rho$ and $\Theta$ are, respectively, the prior probability density and the probability density of the theoretical model on $(\dv,\mv)$. The function $\rho$ contains the information on the system (parameter and data) before the inference process. This can be either the observational data or the prior information on the parameters of the model. $\Theta$ represents the uncertainties on the theoretical model. 

We now switch back to the notations of Sect.~\ref{sect:methodo} using the relations ${\bf d} = \{\Omega_i\}$ and ${\bf m} = \{\theta_i\}$. The data and the parameters are independent, hence, we can write $\mu(\Omega_i,\theta_i) = \mu_{\Omega_i}(\Omega_i)\mu_{\theta_i}(\theta_i)$. In this work we choose $\mu_{\Omega_i}$ and $\mu_{\theta_i}$ as uniform densities \citep{Tarantola04}. Using the same argument of independence, we can write the prior probability density as a product of probability densities $\rho(\Omega,\theta) = \rho_{\Omega_i}(\Omega_i)\rho_{\theta_i}(\theta_i)$, where $\rho_{\Omega_i}$ is obtained from the Gaussian-process modelling. $\rho_{\theta_i}$ is uniform on the latitude interval $[0,90^{\circ}]$, since we have no prior information on the latitude of the active region. The range chosen is the consequence of our lack of spatial resolution between the two stellar hemispheres as mentioned in Sect.~\ref{sect:result}. Finally we write $\Theta(\Omega_i,\theta_i) = \eta_{\Omega_i}(\Omega_i|\theta_i)\mu_{\theta_i}(\theta_i)$ with $\eta_{\Omega_i}(.|\theta_i)$ the probability density on $\Omega_i$ conditional on $\theta = \theta_i$, i.e. the probability density of the theoretical rotation rate taken at $\theta_i$ (see the central panel of Fig.~2 for an illustration). These choices ensure that $\sigma$ is normalizable and can be treated as a probability density.

The final step, once $\sigma(\Omega_i,\theta_i)$ has been estimated, is to obtain the posterior probability density on $\theta_i$ only instead on $(\Omega_i,\theta_i)$. This is achieved by marginalizing the rotation rate, i.e. performing the integral
\begin{equation}
\displaystyle
\sigma_{\theta_i}(\theta_i) = \int \sigma(\Omega_i,\theta_i)d\Omega_i,
\end{equation}
over the relevant range of rotation rates. The density $\sigma_{\theta_i}$ is called the marginal probability density for $\theta_i$. The $\sigma_{\theta_i}$ are the objects represented in Fig.~3 as a function of time $t$ and $\theta_i$ and is what we call the butterfly diagram.

\end{document}